\documentclass[aps,prl,amsmath,amssymb,twocolumn,showpacs,showkeys,superscriptaddress, floatfix]{revtex4-1}

\usepackage{graphicx}% Include figure files
\usepackage{dcolumn}% Align table columns on decimal pointuse
\usepackage{color}
\usepackage{float}
\usepackage{multirow}
\usepackage{hyperref}

\begin{document}
%\title{Carrier dynamics of self-assembled (InGa)(AsSb)/GaAs quantum dots on GaP(001)}

%\title{Importance of antimony for temporal evolution of emission from self-assembled (InGa)(AsSb)/GaAs quantum dots on GaP(001)}

\title{On the importance of antimony for temporal evolution of emission from self-assembled (InGa)(AsSb)/GaAs quantum dots on GaP(001)}

%\title{Antimony modulated temporal evolution of emission from self-assembled (InGa)(AsSb)/GaAs quantum dots on GaP(001)}

\author{Petr Steindl}
\email[]{steindl@physics.leidenuniv.nl}
\affiliation{Department of Condensed Matter Physics, Faculty of Science, Masaryk University, Kotl\'a\v{r}sk\'a~267/2, 61137~Brno, Czech~Republic}
%\affiliation{Central European Institute of Technology, Masaryk University, Kamenice 753/5, 62500~Brno, Czech~Republic}
\affiliation{Huygens-Kamerlingh Onnes Laboratory, Leiden University, P.O. Box 9504, 2300 RA Leiden, Netherlands}
% if you have orcid, write here, please
% https://orcid.org/0000-0001-9059-9202

\author{Elisa Maddalena Sala}
\email[]{e.m.sala@sheffield.ac.uk}
\affiliation{Center for Nanophotonics, Institute for Solid State Physics, Technische Universit\"{a}t Berlin, Hardenbergstr. 36, 10623 Berlin, Germany}
%\affiliation{Technische Universit\"{a}t Berlin, Institut f\"{u}r Festk\"{o}rperphysik, Hardenbergstr. 36, 10623 Berlin, Germany}
\affiliation{EPSRC National Epitaxy Facility, The University of Sheffield, North Campus, Broad Lane, S3 7HQ Sheffield, United Kingdom}
% if you have orcid, write here, please

\author{Benito Al\'{e}n}
\affiliation{Instituto de Micro y Nanotecnolog\'{i}a, IMN-CNM, CSIC (CEI UAM+CSIC) Isaac Newton, 8, E-28760, Tres Cantos, Madrid, Spain}
% if you have orcid, write here, please
% https://orcid.org/0000-0003-3939-1611

\author{Dieter Bimberg}
\affiliation{Center for Nanophotonics, Institute for Solid State Physics, Technische Universit\"{a}t Berlin, Germany}
\affiliation{``Bimberg Chinese-German Center for Green Photonics'' of the Chinese Academy of Sciences at CIOMP, 13033 Changchun, China}

\author{Petr Klenovsk\'y}
\email[]{klenovsky@physics.muni.cz}
\affiliation{Department of Condensed Matter Physics, Faculty of Science, Masaryk University, Kotl\'a\v{r}sk\'a~267/2, 61137~Brno, Czech~Republic}
%\affiliation{Central European Institute of Technology, Masaryk University, Kamenice 753/5, 62500~Brno, Czech~Republic}
\affiliation{Czech Metrology Institute, Okru\v{z}n\'i 31, 63800~Brno, Czech~Republic}
% if you have orcid, write here, please

\date{\today}

\begin{abstract}
Understanding the carrier dynamics of nanostructures is the key for development and optimization of novel semiconductor nano-devices. Here, we study the optical properties and carrier dynamics of (InGa)(AsSb)/GaAs/GaP quantum dots (QDs) by means of non-resonant energy and temperature modulated time-resolved photoluminescence. Studying this material system is important in view of the ongoing implementation of such QDs for nano memory devices. Our set of structures contains a single QD layer, QDs overgrown by a GaSb capping layer, and solely a GaAs quantum well, respectively. Theoretical analytical models allow us to discern the common spectral features around the emission energy of 1.8~eV related to GaAs quantum well and GaP substrate. We observe type-I emission from QDs with recombination times between 2~ns and 10~ns, increasing towards lower energies. The distribution suggests the coexistence of momentum direct and indirect QD transitions. Moreover, based on the considerable tunability of the dots depending on Sb incorporation, we suggest their utilization as quantum photonic sources embedded in complementary metal-oxide-semiconductor (CMOS) platforms, since GaP is almost lattice-matched to Si. Finally, our analysis confirms the nature of the pumping power blue-shift of emission originating from the charged-background induced changes of the wavefunction topology.
\\

\end{abstract}

%\pacs{73.21.La, 75.75.-c, 85.35.Be, 68.65.Hb}
%

\maketitle

\section{Introduction}

In the last few decades, nano-structures like self-assembled III-V QDs have been investigated due to their wide range of novel physical properties. Advantages in this respect led to a number of different applications, such as active media in semiconductor lasers~\cite{Bimberg1997,Ledentsov,Heinrichsdorff1997}, as building blocks for quantum information devices, particularly for quantum repeaters~\cite{Bimberg2008_EL,Azuma_Qrep,Li2019}, as efficient single and entangled photon sources~\cite{Lochamnn2006,muller_quantum_2018,martin-sanchez_single_2009, schlehahn_single-photon_2015,paul_single-photon_2017,salter_entangled-light-emitting_2010,Aberl:17,Klenovsky2018, Senellart2017}, including highly-entangled  states for quantum computing~\cite{Lim_PRL2005, Lindner_PRL2009,Istrati2020, steindl2020artificial}, or as nanomemories~\cite{Marent2011,BimbergPatent,Marent2009_microelectronics,Bimberg2011_SbQDFlash, Marent_APL2007_10y}. Among III-V QDs, particularly type-I indirect (InGa)(AsSb)/GaAs QDs embedded in a GaP(001) matrix~\cite{t_sala,Sala2018} have recently attracted attention due to their promising use as storage units for the QD-Flash nanomemory cells~\cite{t_sala,Sala2018}, as potentially effective entangled photon sources~\cite{Klenovsky2018_TUB}, owing to their smaller fine-structure splitting (FSS) of the ground state exciton compared to well-known type-I systems such as (InGa)As/GaAs~\cite{Aberl:17,Klenovsky2018}, and as quantum gates~\cite{Burkard_PRB1999_QuantumGate,Krapek2010,Klenovsky2016,Klenovsky2018_TUB}. The concept of hole storage QD-Flash was initially suggested by Bimberg and coworkers~\cite{Marent2011,BimbergPatent,Marent2009_microelectronics,Bimberg2011_SbQDFlash, Marent_APL2007_10y,Kapteyn1999} following first pioneering studies~\cite{Kapteyn1999} regarding the mechanisms of electron escape from InAs/GaAs QDs, by using the Deep Level Transient Spectroscopy (DLTS).
The key feature of the QD-Flash is to combine the fast access times of Dynamic Random Access Memories (DRAM) with the non-volatility of the Flash, which leads to a universal memory type, potentially simplifying future computer architectures. Recently, type-I indirect (InGa)(AsSb)/GaAs/GaP QDs showed an improvement of one order of magnitude in the storage time compared to pure In$_{0.5}$Ga$_{0.5}$As/GaAs/GaP QDs~\cite{Bonato_APL2015,Stracke2014}, reaching $\sim$1 hr at room temperature~\cite{t_sala,Sala2018}. This result represents to date the record for Metal-Organic Vapor Phase Epitaxy (MOVPE)-grown QDs, thus opening up the possibility to use this technique to fabricate memory devices based on high-quality III-V semiconductor QDs. Additionally, in Ref.~\cite{Klenovsky2018_TUB} the authors theoretically discussed the physical properties of such material system -- particularly the quantum confinement type -- depending on the relative In/Ga and As/Sb contents in the QDs. It was found that these QDs showed concurrently both direct and indirect optical transitions for increasing Sb content, finally leading to type-II band alignment~\cite{Klenovsky2018_TUB}. That made such QDs be excellent candidates for quantum information technologies. Increasing the Sb content in the QDs has been previously made possible by overgrowing (InGa)(AsSb)/GaAs/GaP QDs with a GaSb capping layer, which has effectively modified the QD composition~\cite{Steindl2019_PL}. Moreover, through detailed investigations of their optical properties, it was found that such procedure led to an energy swapping of the $\Gamma$ and L states, thereby increasing the wavefunction leakage outside the QDs~\cite{Klenovsky2018_TUB,Steindl2019_PL}. This property is indeed very appealing for further improvement of storage times since an increased Sb incorporation into the QDs leads to increased hole localization energy~\cite{Klenovsky2018_TUB,Bimberg2011_SbQDFlash,Marent_APL2007_10y}. Finally, fabricating QDs on GaP substrates is advantageous in terms of integration on Silicon platforms, since the lattice mismatch between GaP and Si amounts to just 0.4\%, thus making defect-free MOVPE growth of GaP on Si possible~\cite{Grassman_apl2013}. 

In this work, we take the next step and study the carrier dynamics of (InGa)(AsSb)/GaAs/GaP QDs, by means of time-resolved-photoluminescence (TRPL) for varying detection energy and sample temperature. This allows us to energetically separate the overlapping optical transitions previously observed in our recent work~\cite{Steindl2019_PL}. First, we provide a brief overview of our sample structures. Afterwards, we discuss the experimental results on carrier lifetimes for varying measurement conditions. Analytical models, describing the observed physical phenomena are provided, leading us to discern the different types of optical transitions involved. We would like to point out that, to date, there is no such detailed optical investigation of this material system.
%%%%%%%%%%%%%%%%%%%%%%%%%%%%%%%%%%%%%%%%
% STRUCTURAL ANALYSIS
%%%%%%%%%%%%%%%%%%%%%%%%%%%%%%%%%%%%%%%%
\section{Sample structures}
 The samples were grown by MOVPE in Stranski-Krastanov (SK) mode on GaP(001) substrates at the TU Berlin~\cite{t_sala,Sala2018}. Such samples were also previously investigated by means of steady-state photoluminescence~\cite{Steindl2019_PL}. The structures of the samples studied in this work are schematically depicted in all figures as insets. 

All samples include 5~ML-thick GaAs interlayer (IL), a crucial ingredient for the subsequent QD formation, as pointed out by Sala~{\sl et al.}~\cite{t_sala,Sala2016}. The sample having the IL only is referred to as S$_\mathrm{w/o}$, that labeled S$_\mathrm{with}$ (S$_\mathrm{cap}$) contains (InGa)(AsSb) QDs, without (with) $\sim$1~ML GaSb capping. The QDs are of truncated pyramid shape, with basis diameter of $\sim$15~nm and height of $\sim$2.5~nm~\cite{Klenovsky2018_TUB,Steindl2019_PL, Gajjela2020}. For detailed information about the growth procedure, see~\cite{Sala2018,t_sala,Steindl2019_PL}. Additional details on their structure, particularly on size, shape, and composition, can be found in very recent work on XSTM and atom probe tomography investigations on such QD samples~\cite{Gajjela2020}. 

The sample photoluminescence (PL) is found at $\sim$1.8~eV and shows several not well spectrally separated bands, representing a combination of momentum direct and indirect type-I transitions from QDs~\cite{Steindl2019_PL}. 
%%%%%%%%%%%%%%%%%%%%%%%%%%%%%%%%%%%%%%%%
% SET UP
%%%%%%%%%%%%%%%%%%%%%%%%%%%%%%%%%%%%%%%%
\section{Experimental setup for TRPL measurements}
In TRPL experiments we used a pulsed laser with the wavelength of 405~nm, focused on 0.06~mm$^2$ area with a 60~ps pulse-width. The emitted PL spectrum was dispersed by 1200~grooves/mm ruled grating and detected by a Si avalanche photodiode (APD). First, we cooled the samples to 15 K, and detected in 200~ns temporal window the energy-resolved TRPL signal for each wavelength. Then, within temperature-resolved TRPL, the sample temperature $T$ was varied in the range 15--130$\,$K. Here, the temporal window was modified to maximize the resolution from 200$\,$ns for lower $T$, to 25$\,$ns for higher $T$. Changing the temporal window is connected with changes in repetition rate, which was varied between 5$\,$MHz (for the temporal window 200$\,$ns; used also for energy-resolved TRPL) and 80$\,$MHz (for 25$\,$ns).
%%%%%%%%%%%%%%%%%%%%%%%%%%%%%%%%%%%%%%%%
% LINE SHAPE MODEL
%%%%%%%%%%%%%%%%%%%%%%%%%%%%%%%%%%%%%%%%
\vspace{1cm}
\section{Spectral line-shape model}

\begin{figure*}
	\centering
	\includegraphics[width=1\linewidth]{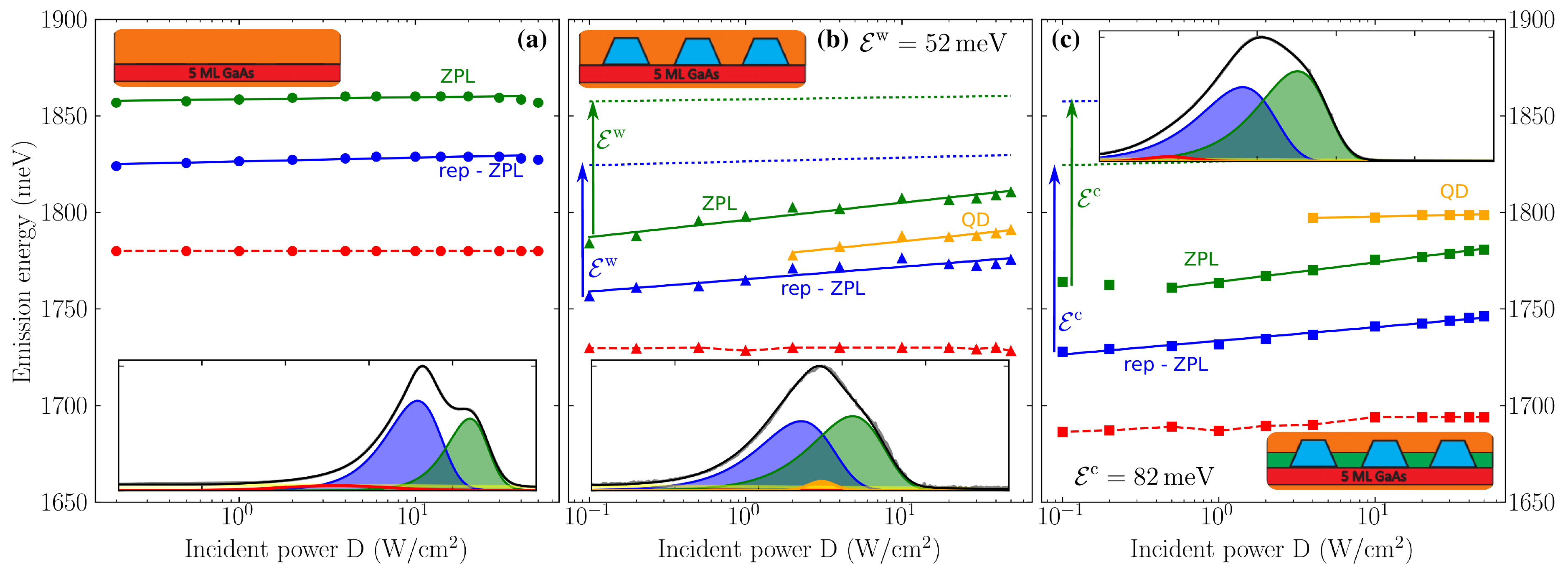}
	\caption{ Excitation power dependence of emission energies of samples (a) $\mathrm{S}_\mathrm{w/o}$, (b) $\mathrm{S}_\mathrm{with}$, and (c) $\mathrm{S}_\mathrm{cap}$. Symbols represent the emission energies fitted from PL spectra. A typical (normalized) spectrum of each sample measured with $D=3.3$~W/cm$^2$ together with colored band-reconstruction over spectral range of 1650--1900~meV is shown in insets. The emission energies evolve in agreement with diffuse interface model for spatial type-I transitions~\cite{Abramkin_blueshift_analytical, Steindl2019_PL} (solid lines). Low-power emission energies of IL transitions in $\mathrm{S}_\mathrm{with}$ ($\mathrm{S}_\mathrm{cap}$) are red-shifted by $\mathcal{E}^\mathrm{w}$ ($\mathcal{E}^\mathrm{c}$) in respect to that in $\mathrm{S}_\mathrm{w/o}$.} 	\label{fig:All_E_int}
\end{figure*}

For the description of PL in the time domain (TDPL), we take advantage of the similarity in the grown structures, leading to expected shared spectral features across samples associated with carriers confined in the GaAs IL,~i.e., zero-phonon (ZPL) and phonon-replica (rep-ZPL) transitions of electrons from $X_{xy}$ conduction minima to $\Gamma$ valence band maximum~\cite{Prieto_APL1997, Steindl2019_PL}. Through analysis of the line-shape in the $S_\mathrm{w/o}$ sample, we conclude that the convolution of two asymmetrical bands with maximum emission energy $E_\mathrm{max}$ concurrently showing a small high-energy and a prominent low-energy band-tail produce better results than the purely Gaussian spectral deconvolution used in Ref.~\cite{Steindl2019_PL}. 
The low energy tail shall be related to carrier localization into long-range IL potential fluctuations~\cite{Almosni2016}. Meanwhile, high energy tails shall be related to phonon-assisted thermal population of delocalized states, especially at large excitation powers/temperatures or during the initial stages of the relaxation process. We follow the work of Almosni~\textit{et al.} to describe the low energy tail long-range fluctuations through the following  equation~\cite{Almosni2016}

%thermalization~\cite{Amtout1995} and the long-range IL fluctuations/disorder~\cite{Almosni2016}. Therefore, we move from purely statistical spectral-shape of these bands discussed in Ref.~\cite{Steindl2019_PL} to line shape $I_\mathrm{IL}$, which combines the above-mentioned effects. 
%Here, the long-range fluctuations are taken into the line-shape model~\cite{Almosni2016}
%
\begin{eqnarray}
I \propto \frac{\exp(\epsilon/E_\mathrm{long})}{E_\mathrm{long}} \exp(-\exp(\epsilon/E_\mathrm{long})) \label{eq:I_long}
\end{eqnarray}
where a single parameter $E_\mathrm{long}$ characterizes the long-range potential disorder energy. Meanwhile, hot carrier population is taken into account through an $n$ phonon-assisted thermalization process by line-shape~\cite{Amtout1995} 
\begin{eqnarray}
I_n\propto \epsilon^{5/2-n}\exp\left(-\frac{\epsilon}{k_B T_\mathrm{ca}}\right) \label{eq:I_rep}
\end{eqnarray}
with carrier thermalization energy of $k_BT_\mathrm{ca}$; $\epsilon=E-E_\mathrm{max}$. We limit our description of $I_\mathrm{IL}$ (convolution of Eqs.~(\ref{eq:I_long}) and (\ref{eq:I_rep})) to one-photon process ($n=1$) only.

As it can be seen in Fig.~\ref{fig:All_E_int}, two replicas of the above lineshape model account for most of the PL emission in these samples, yet not completely. To describe the full PL spectrum, two additional Gaussian profiles are necessary. One of them describes a rather broad band (FWHM larger than 35~meV), clearly observable only at very low excitation powers, likely originating in the donor-acceptor pair (DAP) transitions in GaP~\cite{Dean_PR68,Dean_1970} or other defect induced during GaAs IL and QDs formation (the latter in the case of samples with QDs). We attribute the second Gaussian band to the recombination from QDs, being due to non-optimized excitation wavelength, and thus very weak and observable mainly for high excitation powers. Before moving to time-resolved analysis, we show the validity of the fitting model by applying it to the PL vs. continuous-wave excitation power dependence $D$ measured at 15~K and published in our previous study~\cite{Steindl2019_PL}. 

Similarly as there, the fitted peak energies are used to analyse the emission blue-shift with increasing $D$, in order to determine the type of carrier spatial confinement. Although elsewhere in the literature~\cite{Klenovsky2017,Jo2012,Ledentsov1995,Jin,Gradkowski_pssb2009} the presence of blue-shift is automatically assigned to indirect spatial alignment, the so-called type-II, we examine here the blue-shift by $E=E_0+U\mathrm{ln}(D) + \beta D^{1/3}$~\cite{Abramkin_blueshift_analytical,Steindl2019_PL} allowing us to disentangle type-II bend-bending, due to state squeezing represented by the parameter $\beta$, from the spatial alignment independent blue-shift caused by crystalline defects described by the Urbach energy tail $U$. 
Having $\beta$ negligible, the analysis in Fig.~\ref{fig:All_E_int} suggests that the emission bands of our heterostructures are of type-I,~i.e. spatially direct, as also previously reported based on Gaussian fits~\cite{Steindl2019_PL} and in agreement with $\mathbf{k\cdot p}$ simulations~\cite{Klenovsky2018_TUB}. Moreover, we observe that ZPL and rep-ZPL transitions of samples $\mathrm{S}_\mathrm{with}$ and $\mathrm{S}_\mathrm{cap}$ are red-shifted in respect to their energies observed from PL of $\mathrm{S}_\mathrm{w/o}$ by $\mathcal{E}^\mathrm{w}=52$~meV and $\mathcal{E}^\mathrm{c}=82$~meV, respectively. This shift partially reflects the strain-relaxation initialized by constituent segregation from QD-layer~\cite{Gajjela2020} and, thus, partially induced change in band confinement. The former is connected also with the natural spectral broadening when additional localized defect states are created in the heterostructure. These additional states then form an effective background potential increasing with excitation power, leading to the energy blue-shift of bands of samples with QDs, characterized by the Urbach energy. However, the bands of the sample with only GaAs IL do not manifest themselves. 
A similar shift can be also observed in the time domain after the non-resonant pulse-excitation when the carriers first thermalize into the trap states and form the initial background potential. As those recombine, $E_\mathrm{long}$ decreases, the potential weakens and, thus, the emission energy is gradually red-shifted, as we will discuss later in more detail. This potential weakening is connected also with the spreading of the state wavefunctions, effectively observable as an increase in recombination times in the excitation resolved TRPL, see supplemental information~\cite{Supplement}.

Although we attribute the QD band in the emission of samples with dots, we expect in the studied spectral range even richer spectral response related to momentum-indirect transitions of QDs~\cite{Klenovsky2018_TUB} and their compositional variations~\cite{Gajjela2020} which are most likely shadowed by much stronger GaAs IL emission.

\begin{center}
\begin{table*}%[ht]
{\small
\hfill{}
\caption{Summary of the best-fit parameters of the spectral shape model applied to the excitation power resolved PL and TDPL of all studied samples. Symbol $^*$ ($^{**}$) refers to a discrepancy of $+10$ meV ($-5$ meV) in $E_0$ from TDPL in respect to the extracted value from the excitation power-dependent PL. For ZPL and rep-ZPL, we give $E_\mathrm{long}$ as FWHM.}
\begin{ruledtabular}
\begin{tabular}{llccccccc}
	sample & transition &  FWHM (meV)&  $E_\mathrm{0}$ (meV)& $U$ (meV)&  $\Delta E$ (meV)  & $\tau_E$ (ns)& $\tau_1^\mathrm{TDPL}$ (ns)&  $\tau_2^\mathrm{TDPL}$ (ns)  \\ 
\hline 
    \multirow{2}{*}{$\mathrm{S}_\mathrm{w/o}$ } 	&	ZPL &  10 &  $1858\pm0.4$& $0.5\pm0.2$&  $1.3\pm0.4$& $50\pm40$&  $10.7\pm0.2$  & $52\pm1$\\ 
        &	rep-ZPL &  14&  $1826\pm0.4$& $0.8\pm0.1$ & $5.9\pm0.4$  &  $31\pm5$ & $11\pm3$ &$87.6\pm0.7$\\ \hline

	\multirow{3}{*}{$\mathrm{S}_\mathrm{with}$ } & ZPL &  19 &  $1796^*\pm1$ &$3.9\pm 0.4$ & $13.8\pm0.5$  & $41\pm4$& $6.8\pm0.1$ & $47\pm1$ \\ 
	&	rep-ZPL&  20 &  $1765^*\pm 1$ & $2.8\pm0.4$ & $11\pm 1$ &   $46\pm6$&   $12.9\pm0.5$ & $47\pm1$\\ 
    %& QDs &  19 &  $1777^*\pm2$& $3.6\pm0.6$ 		& $10\pm3$ & $30\pm20$&$10.4\pm0.1$ &   \\ \hline
    & QDs &  19 &  $1777^*\pm2$& $3.6\pm0.6$ 		& $14.3\pm0.5$ & $35\pm4$&$10.4\pm0.1$ &   \\ \hline
    
     \multirow{3}{*}{$\mathrm{S}_\mathrm{cap}$ } &      ZPL&  20 &  $1764\pm0.4$ &  $4.4\pm0.1$&  $ 17\pm1$ & $44\pm 7$& $14.9\pm0.1$&  $2.0\pm0.1$ \\ 
    &	rep-ZPL &  23&  $1733\pm 0.4$ & $3.1\pm0.2$ &  $ 5.4\pm 0.7$ & $19\pm4$& $68\pm4$&  \\ 
    %&	QDs &  8 &  $1796^{**}\pm0.6$& $0.7\pm0.2$ & $10\pm2$ &   $3\pm1$&$7.7\pm2$& \\ 
    &	QDs &  8 &  $1796^{**}\pm0.6$& $0.7\pm0.2$ & $10\pm1$ &   $4.1\pm0.4$&$7.7\pm2$& \\ 
\end{tabular} \label{tab:S_wo_TDPL} \label{tab:Sw_QSs_TDPL} \label{tab:Scap_QSs_TDPL}
\end{ruledtabular}}
\hfill{}
{\small
\hfill{}
\caption{Parameters obtained from Gourdon and Lavallard model, Eq.~(\ref{eq:En_TRPL}). Units of the variables are: $\tau_\mathrm{r}^i$ is in ns, $E_\mathrm{me}^i$ and $U_0^i$ are in meV.}
\begin{ruledtabular}
\begin{tabular}{lccc|ccc|ccc|ccc}
\multirow{2}{*}{sample} &\multicolumn{3}{c}{GaAs IL} & \multicolumn{3}{c}{GaAs IL, phonon rep.}& \multicolumn{3}{c}{growth defects}& \multicolumn{3}{c}{DAP in GaP}\\
		 &   $\tau_\mathrm{r}^\mathrm{ZPL}$ & $E_\mathrm{me}^\mathrm{ZPL}$  & $U_\mathrm{0}^\mathrm{ZPL}$  &   $\tau_\mathrm{r}^\mathrm{rep}$ & $E_\mathrm{me}^\mathrm{rep}$  & $U_\mathrm{0}^\mathrm{rep}$   &
		  $\tau_\mathrm{r}^\mathrm{d}$  & $E_\mathrm{me}^\mathrm{d}$  & $U_\mathrm{0}^\mathrm{d}$ & $\tau_\mathrm{r}^\mathrm{DAP}$  & $E_\mathrm{me}^\mathrm{DAP}$  & $U_\mathrm{0}^\mathrm{DAP}$  \\ 
\hline 
         $\mathrm{S}_\mathrm{w/o}$& $13.0\pm1.0$ & $1882\pm 3$ & $4\pm2$ & $14.4\pm2.4$ & $1856\pm 2$ & $4.3\pm1.4$ & $90\pm1$ &  $1877\pm1$ &  $5.3\pm0.2$&  $ 260\pm 30$ & $1776\pm3$ &  $15.6\pm0.5$ \\ 
		$\mathrm{S}_\mathrm{with}$&  $31.5\pm0.7$ &  $1835\pm 1$ & $8.0\pm0.6$ &  $30.7\pm 0.3$& $1801\pm 2$  &$2.7\pm 1.1$ &  $ 284\pm 2$ & $ 1810\pm 1$ & $ 14.9\pm 0.1$ &$561\pm1$ & $1781\pm1$  &$17.0\pm0.1$ \\ 
		$\mathrm{S}_\mathrm{cap}$ &  $18.4\pm 0.5$ &  $1792\pm1$& $11\pm1$ & $18.8\pm0.3$& $1743\pm1$ &$2.5\pm 0.9$  & &  &  & $1156\pm 1$ &  $1737\pm1$& $17.6\pm0.2$
		\\ 
\end{tabular} \label{tab:Energy_TRPL}
\end{ruledtabular}}
\hfill{}
\end{table*}
\end{center}

%%%%%%%%%%%%%%%%%%%%%%%%%%%%%%%%%%%%%%%%
% EMISSION ENERGY DEPENDENT TRPL
%%%%%%%%%%%%%%%%%%%%%%%%%%%%%%%%%%%%%%%%
\section{Emission energy dependent TRPL}

\begin{figure*}[!ht]
	\centering
	\includegraphics[width=1\linewidth]{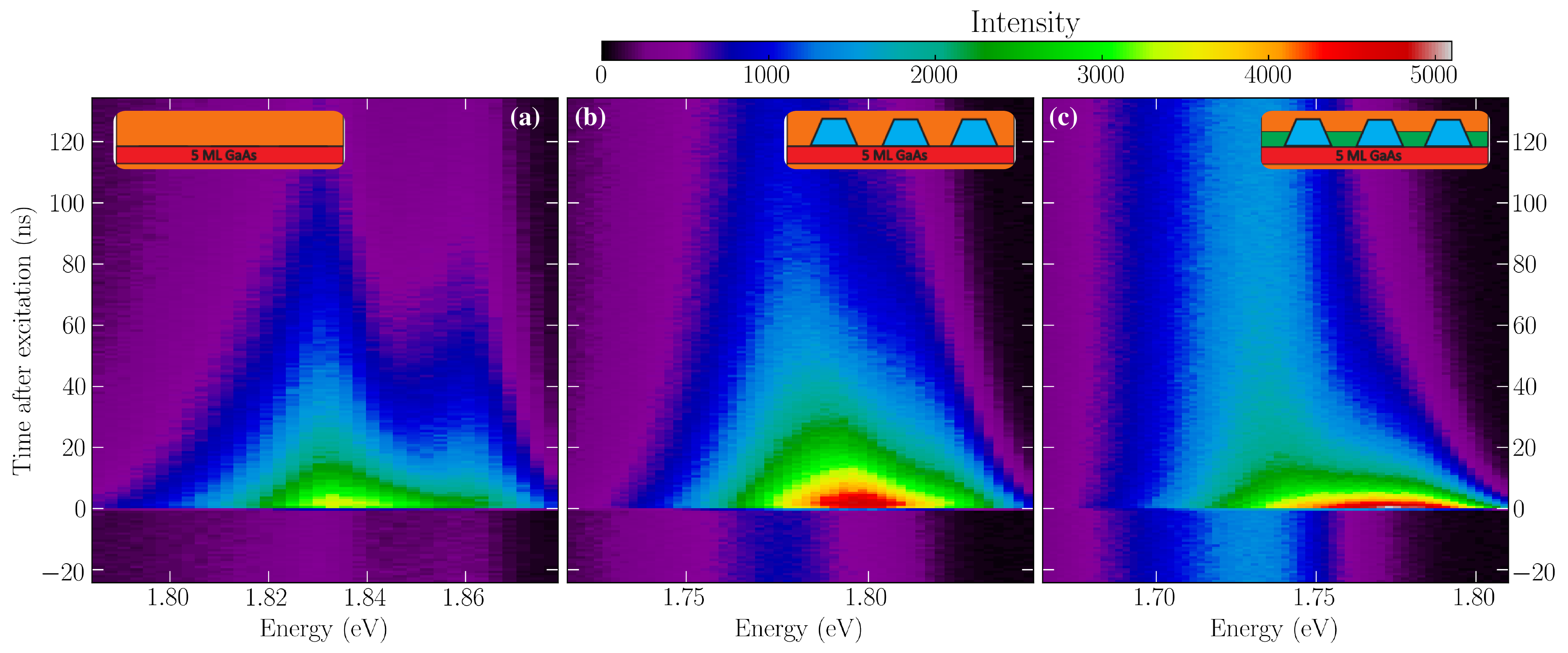}
	\caption{False-color plots of PL intensity as a function of time and emission energy for samples (a) $S_\mathrm{w/o}$, (b) $S_\mathrm{with}$, and (c) $S_\mathrm{cap}$. The color scale is identical for all samples.
	}
	\label{fig:TDPL_map}
\end{figure*}

\begin{figure*}
	\centering
	\includegraphics[width=1\linewidth]{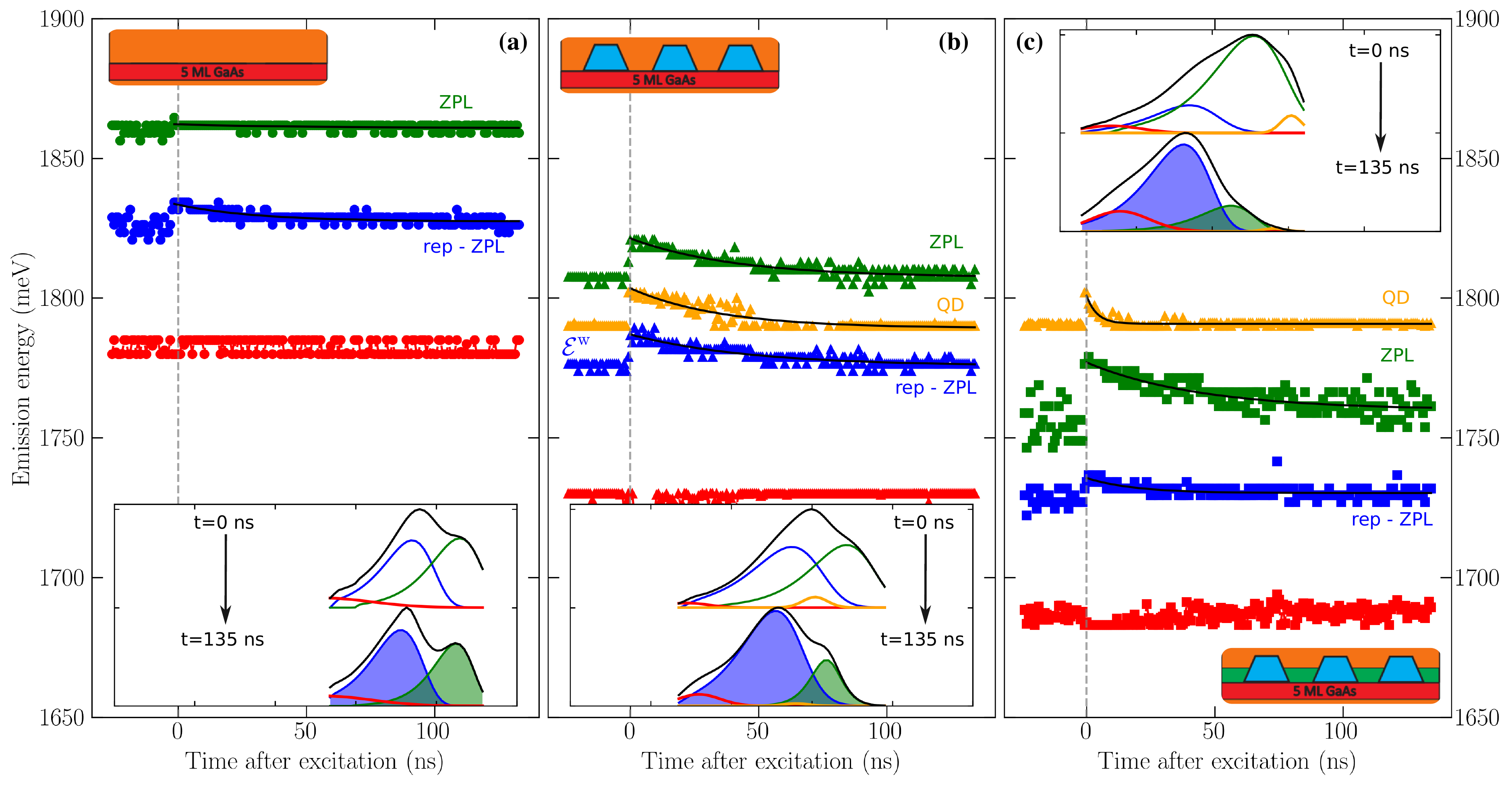}
	\caption{Fitted TDPL emission energies (symbols) which exhibit exponential-like energy red-shift with temporal evolution (fit, black solid lines). While for $\mathrm{S}_\mathrm{w/o}$ in (a), the shift is timid, for samples $\mathrm{S}_\mathrm{with}$ (b) and $\mathrm{S}_\mathrm{cap}$ (c) it exceeds 10~meV and leads to an observable spectral-shape variation within temporal evolution (see Fig.~\ref{fig:TDPL_map} and insets with color-coded fitted emission bands over the spectral range of 1.65--1.9 eV). The broken grey vertical lines indicate the moment of the laser pulse excitation. 
	}
	\label{fig:All_TDPL}
\end{figure*}

\begin{figure*}
	\centering
	\includegraphics[width=1\linewidth]{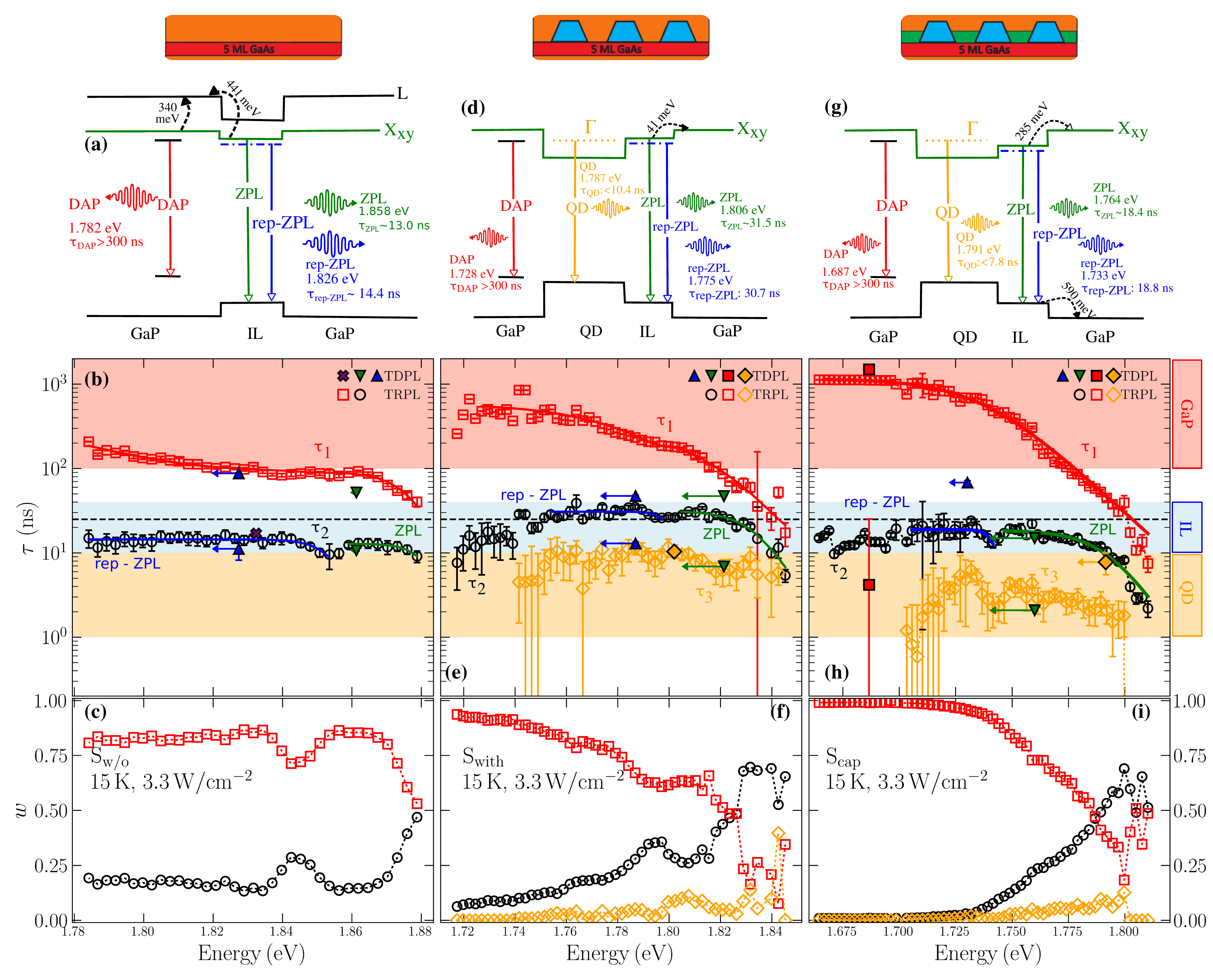}
	\caption{
    Band schemes of samples $\mathrm{S}_\mathrm{w/o}$ [panel (a)], $\mathrm{S}_\mathrm{with}$ [panel (d)] and $\mathrm{S}_\mathrm{cap}$ [panel (g)] according to the observed TDPL transitions $E_0$. The insets show the experimentally observed recombination times, transition (taken from fits of TDPL, solid lines) and escape (dashed line) energies. 
	The energy dispersion of (b) time constants and (c) corresponding weights $w$ for sample $\mathrm{S}_\mathrm{w/o}$ obtained by fitting the TRPL signal by the double mono-exponential model using Eq.~(\ref{eq:I_TRPL}) (symbols) and fitted by the Gourdon-Lavallard's model~\ref{eq:Gourdon} (solid lines)~\cite{Gourdon_PSSB1989}.	That for samples $\mathrm{S}_\mathrm{with}$ and $\mathrm{S}_\mathrm{cap}$ obtained from fitting of the TRPL signal by triple mono-exponential model using Eq.~(\ref{eq:I_TRPL}) is shown in panels (d)--(f) and (g)--(i), respectively.  The deconvoluted time constants show good agreement with TDPL intensity decays (full symbols with arrows representing time-domain $\Delta E$ shift; transitions are assigned by color in agreement with Fig.~\ref{fig:All_TDPL}) and are compared to the recombination time of wetting layer in InAs/GaAs QDs system of 25\,ns (dashed line), taken from Ref.~\cite{Karachinsky_WL25ns_missoriented_substr}. Shaded areas of $1-10$\,ns, $10-40$\,ns, and $>100$\,ns correspond to different recombination channels.
	}
	\label{fig:All_TRPL_energy}
\end{figure*}

In this section, we study the energy-resolved carrier dynamics in our heterostructures by TRPL. To assign the recombination times to the characteristic bands, we first fit the signal (see raw experimental data in Fig.~\ref{fig:TDPL_map}) in individual time bins by the spectral shape model discussed in the previous part, and we refer to this analysis as time-domain PL (TDPL). For the best-fit results presented in Fig.~\ref{fig:All_TDPL}, we use the parameters obtained from steady-state excitation power dependency. Later, we analyse the signal for each wavelength also by the double mono-exponential model (2ME)
\begin{equation}
I(t)=A_1\exp(-t/\tau_1)+A_2\exp(-t/\tau_2), \label{eq:I_TRPL}
\end{equation}
characterized by amplitude $A_1$ ($A_2$) and decay time $\tau_1$ ($\tau_2$) for the slow (fast) decay process. In the case of samples with QDs, we added to the analysis also the third exponential decay component ($\tau_3$), representing the electron-hole recombination in QDs. Finally, we analyze the spectral distribution of the time decay constants $\tau_1$--$\tau_3$ by an analytical model developed by Gourdon and Lavallard~\cite{Gourdon_PSSB1989}:
\begin{eqnarray}
\label{eq:Gourdon}
\tau = \frac{\tau_\mathrm{r}}{1+\exp( (E-E_\mathrm{me})/U_0)} \label{eq:En_TRPL}
\end{eqnarray} 
which is widely used in the literature~\cite{Rubel_APL2007,Sugisaki_PRB2000}, even though in Eq.~(\ref{eq:Gourdon}) the hopping processes~\cite{Gourdon_PSSB1989} or temperature dependence~\cite{Zhicheng_SciRep2017} are not included. The meaning of the parameters in Eq.~(\ref{eq:Gourdon}) is as follows: $\tau_\mathrm{r}$ is the exciton radiative lifetime, $E_\mathrm{me}$ the characteristic energy for which the radiative time equals the transfer one, analogously to a mobility edge~\cite{Oueslati_PRB1988,Sugisaki_PRB2000}, and $U_0$ is the measured energy of localized states, similar to Urbach energy tail, responsible for the observed energy blue-shift~\cite{Abramkin_blueshift_analytical}.
Note, that $\tau_1$ process decays rather slowly and does not completely disappear in one temporal window, therefore we take into account its repumping from previous pulses in TRPL fits, as discussed in the appendix. This issue is overcome in TDPL by disentangling individual transitions by line-shape model fitting, where the slowest decay is assigned to (mainly non-radiative) pair recombination of DAP in GaP~\cite{Dean_PR68,Dean_1970}. Moreover, in spectral dependence for the evaluation of $\tau_1$ we need to extend the model (\ref{eq:En_TRPL}) by an additional contribution, likely connected with other defects created during the epitaxial growth process.

\subsection{Sample without QDs $\mathrm{S}_\mathrm{w/o}$}
We start our discussion with the sample $\mathrm{S}_\mathrm{w/o}$. TDPL deconvolution allows us to study not only the relaxation-time constants of the considered decay process but also the energy changes of the state in the time domain. Specifically, the emptying of the impurity states entails an exponential-like decrease of the emission energies of the total energy $\Delta E$ for both ZPL and rep-ZPL bands, also recently observed for relaxed GaAs/GaP QDs with type-I band-alignment~\cite{Shamirzaev_APL2010}: 
\begin{eqnarray}
E(t)=E_0+\Delta E \exp(-t/\tau_E),\label{eq:E_shift_time}
\end{eqnarray}
where $E_0+\Delta E$ is the energy of the observed state after laser excitation, which exponentially decays proportionally to the time constant $\tau_E$ (an effective time when impurities and defects affect the electron state) to electron energy $E_0$. That can be equally well understood as due to defects at the interfaces between segments of the heterostructure, which create a local electric field (non-equilibrium carriers) leading to red-shift $\Delta E$ of the electron state with energy $E_0$. The carriers then recombine for $\tau_\mathrm{E}$ upon which the eigenvalue of electron state returns to its value without the presence of the local field $E_0$. Note, that the shift $\Delta E$ cannot be caused by inter-valley scattering, which is three orders of magnitude faster than the observed $\tau_E$~\cite{Zollner_APL89}, nor by the thermalization of higher excited states (since $\tau_E>$ radiative recombination times) or thermalization of free-carrier created after excitation which is of one order of magnitude faster, see $T_\mathrm{ca}$ in supplemental information \cite{Supplement}. 

Even though both bands are shifted by few units of meV, similarly to the total blue-shift observed in steady-state experiments, the integral PL spectrum taken at different times of measurement does not show any significant shift and decays equally in time proportionally to the decay around 10-15~ns, see inset of Fig.~\ref{fig:All_TDPL} (a) and table~\ref{tab:S_wo_TDPL}. These values are in good agreement with cryogenic radiative lifetimes of InAs/GaAs wetting layer of 25~ns~\cite{Karachinsky_WL25ns_missoriented_substr}. Note, that since for the studied samples the energy level separations of IL, DAP, and QDs are not clearly distinguishable, we use double mono-exponential decay function (with time constants  $\tau_1^\mathrm{TDPL}$ and  $\tau_2^\mathrm{TDPL}$) to deconvolute the emission intensity, where the origin of the second time constant is assigned according to the following: DAP and other non-radiative defects decay slowly ($\tau_2^\mathrm{TDPL}>40$ ns), whereas quantum dot transition is fast ($\tau_2^\mathrm{TDPL}<10$ ns).

The standard TRPL deconvolution at each wavelength in Fig.~\ref{fig:All_TRPL_energy}~(b) shows two contributions. The faster, being in good agreement with ZPL and rep-ZPL TDPL band decays, with time constants around 13~ns contributes more or less constantly by 20\,\% to the total intensity [panel (c)]. The slower process, related to DAP and crystalline defects, increases the time-constant up to $\sim200$~ns towards lower energies where none transition from GaAs IL is expected~\cite{Klenovsky2018_TUB, Prieto_APL1997} and is saturated below 1.79~eV as expected from the similarity with the two other samples. Note, that similar behaviour with extremely slow (up to few $\mu$s) low-energy transition were independently reported for (In,Ga)As/GaP~\cite{Robert2012,Robert2016}, Ga(As,P)/GaP~\cite{Abramkin_JAP2012}, and GaSb/GaP~\cite{Abramkin2012} as momentum-indirect transitions from QDs. Because we observe such transition not only for our QDs with completely different stoichiometry but also for GaAs/GaP sample clearly without any QDs, we tend to assign the slow transition to defects in GaP substrate~\cite{Jedral1992,Moser1984}, common for all reported structures. 
Furthermore, we note in Fig.~\ref{fig:All_TRPL_energy}(b) a good agreement between TDPL and TRPL time constants, allowing us to deduce, in power and temperature resolved experiments, the character of relaxation based on the results of TRPL measurements only.

\subsection{Sample with QDs $\mathrm{S}_\mathrm{with}$}
The whole spectrum of $\mathrm{S}_\mathrm{with}$ (Fig.~\ref{fig:TDPL_map}), including ZPL and rep-ZPL bands, is also red-shifted in TDPL in respect to that of $\mathrm{S}_\mathrm{w/o}$, approximately by $\mathcal{E}^\mathrm{w}$, see Fig.~\ref{fig:All_TDPL} and table~\ref{tab:Energy_TRPL}. That is close to the energy shift of $E_\mathrm{me}(S_\mathrm{w/o})-E_\mathrm{me}(S_\mathrm{with})=47$~meV for ZPL (55~meV for rep-ZPL) and together with similar time constants $\tau_1^\mathrm{TDPL}$, pointing to similar physics behind the $I_\mathrm{IL}$ transitions. The best fit emission energies of ZPL and rep-ZPL after excitation show non-equilibrium carrier background potential, initially squeezing the electron wavefunction~\cite{Klenovsky2017, llorens_topology_2019}. Later, as the potential weakens, the wavefunction spatially spreads, leading to the gradual red-shift $\Delta E$ of 14~meV and 11~meV for ZPL and rep-ZPL bands, respectively, to their steady-state energies. This time, in agreement with large blue-shift in excitation power-dependent PL, the shifts are more prominent due to significantly increased number of defects created within QD layer formation and later due to additional atom segregation~\cite{Gajjela2020}. 
In addition to the sample $\mathrm{S}_\mathrm{w/o}$, we observe also $\Delta E$ of 14~meV for the TDPL QD band with time constant of $\sim$10~ns, suggesting impurity induced dynamics connected with the GaAs layer. 

The TRPL signal, deconvoluted by Eq.~(\ref{eq:I_TRPL}) by three mono-exponential decay contributions, shows two patterns: one similar to that observed for $\mathrm{S}_\mathrm{w/o}$, and also a much faster one, which we attribute to the emission from QDs. These processes, depicted in panels (d)--(f) of Fig.~\ref{fig:All_TRPL_energy}, have different weight across the measured spectral range. While for energies below 1.75~eV the DAP dynamical processes dominate, they lose importance for larger energies in favor to the processes involving the GaAs IL. The QD contribution is almost negligible in the whole spectral range, except for an increase of $w_3$, corresponding to QDs, centered around 1.80~eV and 1.83~eV, where $w_3$ is larger than 10\%. The mean values of $\tau_3$ in these spectral ranges are $9.0\pm1.0$~ns and $6.0\pm1.0$~ns, respectively. 

For the spectral characteristic of the transitions, the Gourdon and Lavallard model~\cite{Gourdon_PSSB1989} was used by means of one contribution for the process $\tau_2$, and two contributions for the process $\tau_1$. The best-fit values (see Tab.~\ref{tab:Energy_TRPL}) show the mobility edge of the ZPL transition in IL shifted with respect to that of $\mathrm{S}_\mathrm{w/o}$ by 47~meV, which is in the agreement with the shift of the whole spectrum discussed previously. On the other hand, the mobility edge of DAP in GaP remains not affected by the heterostructure. The radiative time of the ZPL (rep-ZPL) band is $31.5\pm0.7$~ns ($30.7\pm0.3$~ns), which is more than two times larger than that of the sample without QDs. That increase can be understood in terms of different material distribution, as an effect of strain relaxation discussed in~\cite{Steindl2019_PL} due to the GaAs IL overgrowth with QDs, leading to the change of the confinement potentials. On the other hand, disorder energies $U_0$ originating from material redistribution -- in our case mainly due to the strain relaxation -- are higher than for $\mathrm{S}_\mathrm{w/o}$, indicating increased disorder of GaAs IL interface, causing not only creation of trap states, but also non-radiative rates at higher energies effectively enlarging the time constants.

\subsection{Sample with GaSb-capped QDs $\mathrm{S}_\mathrm{cap}$}
As previously shown in~\cite{Steindl2019_PL}, overgrowing the QDs with a thin ($\sim$1ML) GaSb cap leads to an effective increase of the Sb content in QDs. Through the TDPL analysis of sample $\mathrm{S}_\mathrm{cap}$ using the line-shape model with emission energies and FWHM adopted from excitation power dependence, we refine the character of the emission band and assign in Fig.~\ref{fig:All_TRPL_energy} the lifetimes of the observed optical transitions, see particularly the fit in inset of Fig.~\ref{fig:All_TDPL}~(c). 

Across the studied spectral range, we again observe similar signatures as in $\mathrm{S}_\mathrm{w/o}$, but red-shifted by $\mathcal{E}^\mathrm{c}$. This shift is also apparent from the comparison of mobility edges subtracted from the Gourdon and Lavallard model~\cite{Gourdon_PSSB1989}, given in Tab.~\ref{tab:Energy_TRPL}. In contrast to the previous samples, we observe also 40~meV shift of DAP mobility edge which is a rather significant change to be caused by a different character of the DAP process only (i.e. type, or concentration) and possibly causing much longer rep-ZPL transition time as extracted from TDPL. However, we do not observe any change of the mobility edge for samples $\mathrm{S}_\mathrm{w/o}$ and $\mathrm{S}_\mathrm{with}$: this might be still connected to the effect of layer-overgrowth on dynamics. On the other hand, we observe almost unchanged ZPL radiative time of $16.2\pm0.2$~ns (and $14.9\pm0.1$~ns from TDPL).

The whole emission spectrum in Figs.~\ref{fig:TDPL_map}(c) and \ref{fig:All_TDPL}(c) shows changes in the shape of emission bands in the time domain, including observable spectrum red-shift. From TRPL deconvolution by three mono-exponential decay curves, it can be seen that the spectrum consists of the fast component at energies greater than 1.75~eV, which completely disappears during the first 50~ns after excitation, and it is rapidly red-shifted during that period. After 50~ns, only a part of the band at energies below 1.75~eV remains bright. In agreement with the observations for $\mathrm{S}_\mathrm{with}$, below 1.74~eV the DAP dynamical processes clearly dominate and their time constant is $\sim$1\,$\mu$s. For larger energies, the emission due to DAP loses importance in favor of GaAs IL processes. For energies larger than 1.76~eV, also the contribution of QDs starts to be noticeable with $w_3$ $\sim$10~\% and $\tau_3$ of 2--6~ns.

%The contribution centered at 1.690\,eV does not shift in energy during our measurements, and it probably originates only from DAP recombination in GaP. 
The time-evolution of the best-fit emission energies of individual transitions from the TDPL fit given in Fig.~\ref{fig:All_TDPL}(c) shows that ZPL and rep-ZPL bands are exponentially red-shifted by 17~meV and 5~meV, respectively, with time constant $\tau_E$ being 19--44~ns. 

The previous analysis showed an increase of QD recombination times with decreasing energy from 6~ns to 9~ns for $\mathrm{S}_\mathrm{with}$, of 1.83~eV and 1.80~eV, respectively, and from 2~ns to 6~ns for $\mathrm{S}_\mathrm{cap}$ of energies close to 1.79~eV and 1.73~eV. The slower recombination times might be assigned to indirect momentum transitions, even though, without detailed single dot spectroscopic study~\cite{Rauter_indirectQD}, this is rather speculative because it could be as well caused by ensemble averaging~\cite{Schimpf2019}. 

%%%%%%%%%%%%%%%%%%%%%%%%%%%%%%%%%%%%%%%%
% TEMPERATURE DEPENDENT TRPL
%%%%%%%%%%%%%%%%%%%%%%%%%%%%%%%%%%%%%%%%
\section{Temperature dependent TRPL}
\begin{figure*}[!h]
	\centering
	\includegraphics[width=0.9\linewidth]{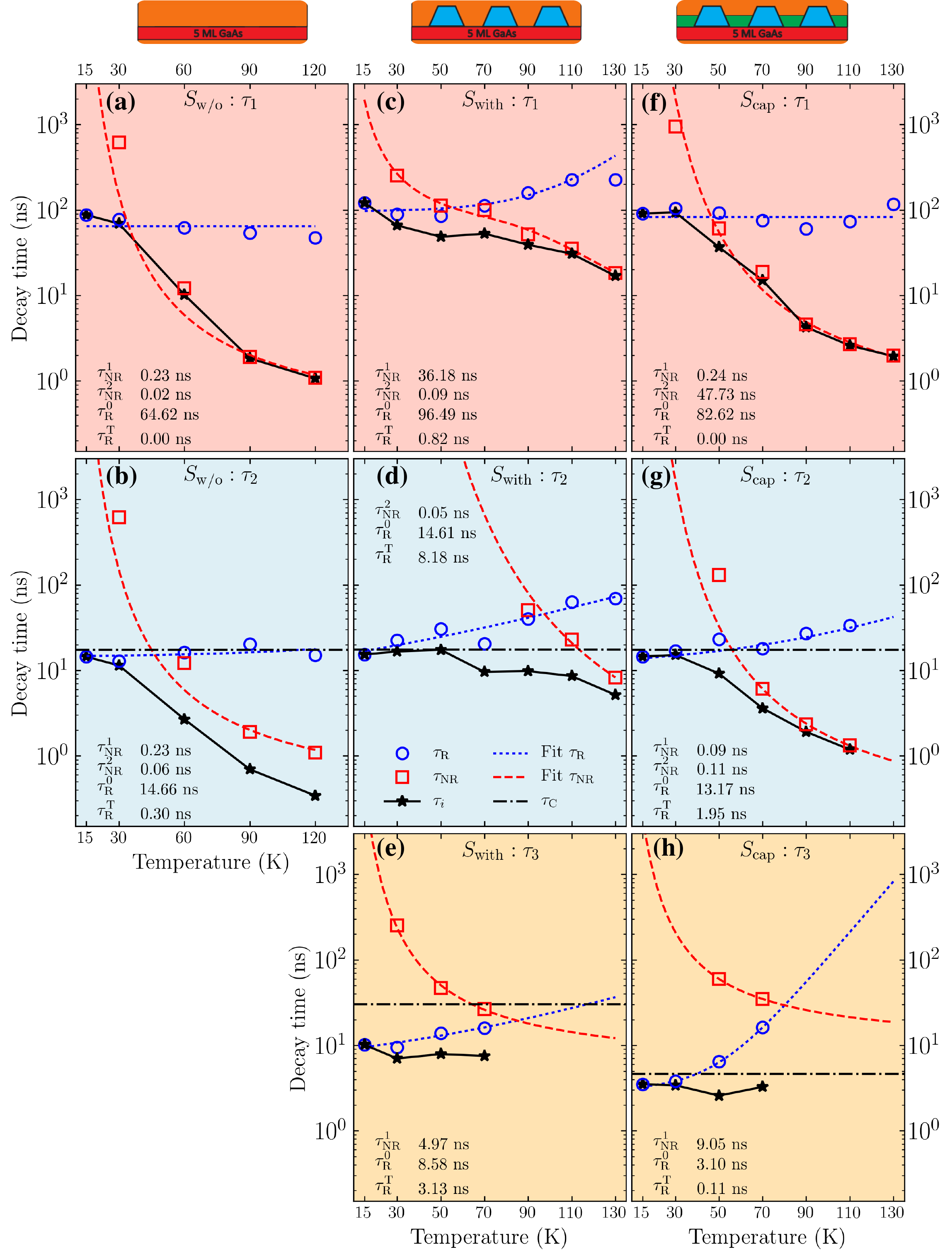}
	\caption{Individual TRPL decay times $\tau_1$--$\tau_3$ (black stars) shown as a function of temperature with the radiative (blue) and non-radiative (red) components for all three samples - panels (a) and (b) show decay times for sample $\mathrm{S}_\mathrm{w/o}$, (c)--(e) that for $\mathrm{S}_\mathrm{with}$ and (f)--(h) for $\mathrm{S}_\mathrm{cap}$. The radiative and non-radiative component (circles and squares) are fitted by Eq.~(\ref{eq:tau_R}) and Eq.~(\ref{eq:nonradiative}) (broken curves), respectively. The best-fit parameters from the models, including $\tau_\mathrm{C}$ (horizontal dash-dot lines), are added for easier comparison.  }
	\label{fig:Temp_TRPL}
\end{figure*}
In this section, we separated radiative and non-radiative contributions of the observed decay times and complete the band schemes in Fig.~\ref{fig:All_TRPL_energy} of the non-radiative processes. Individual recombination channels as a function of $T$ were extracted again using the 3ME (2ME) model for deconvolution of TRPL signal of $\mathrm{S}_\mathrm{with}$ and $\mathrm{S}_\mathrm{cap}$ ($\mathrm{S}_\mathrm{w/o}$). Contrary to the sample $\mathrm{S}_\mathrm{w/o}$, the lifetime of ZPL ($\tau_2$) for samples with QDs ($\mathrm{S}_\mathrm{with}$ and $\mathrm{S}_\mathrm{cap}$) increases with $T$ between 30 and 50~K and thereafter progressively reduces, which is characteristic for the activation of thermally activated escape paths of shallow defects~\citep{Manna_apl2012_TRPLtype2}. Those are most likely generated at the IL/QDs interface during the strain-relaxation caused by QDs overgrowth~\cite{Steindl2019_PL}.

To separate the radiative ($\tau_\mathrm{R}$) and non-radiative ($\tau_\mathrm{NR}$) lifetimes from individual transition channels, we assume, in accordance with
Ref.~\citep{t_alvarez}, that for 15~K the only loss mechanism is the radiative recombination. Thereafter, $\tau_\mathrm{R}$ and $\tau_\mathrm{NR}$ decay times can be extracted from the slow decay time $\tau_1$ by
\begin{equation}
\tau_\mathrm{R}=\frac{I_0}{I_\mathrm{PL}(T)}\tau_1 \label{eq:tau_R_fromtau1},
\end{equation}
and
\begin{equation}
\frac{1}{\tau_1}=\frac{1}{\tau_\mathrm{R}} + \frac{1}{\tau_\mathrm{NR}},
\end{equation}
\noindent where $I_0$ and $I_\mathrm{PL}$ are the PL intensities at 15$\,$K and at larger $T$, respectively. As can be seen in Fig.~\ref{fig:Temp_TRPL}, thermally activated scattering processes cause an exponential decrease characterized by $\tau_\mathrm{NR}$ of localized carriers with $T$. That process can be quantitatively interpreted by the model involving two non-radiative processes
\begin{equation}
\frac{1}{\tau_\mathrm{NR}}=\frac{1}{\tau_\mathrm{NR}^1}\exp{\left(\frac{-E_1}{k_\mathrm{B}T}\right)} + \frac{1}{\tau_\mathrm{NR}^2}\exp{\left(\frac{-E_2}{k_\mathrm{B}T}\right)}, \label{eq:nonradiative}
\end{equation}
characterised by the activation energies $E_1$ and $E_2$ and time constants $\tau_\mathrm{NR}^1$ and $\tau_\mathrm{NR}^2$, respectively.
Conversely, $\tau_\mathrm{R}$ of exciton increases exponentially with $T$
\begin{equation}
\tau_\mathrm{R} = \tau_\mathrm{R}^0 + \tau_\mathrm{R}^T \left[ \exp{\left(\frac{T}{T_C} \right)}- 1 \right]\,, \label{eq:tau_R} 
\end{equation}
where $\tau_\mathrm{R}^0$ ($ \tau_\mathrm{R}^T$) describes the $T$ independent (dependent) part of the radiative lifetime, and $T_C$ is the characteristic value of $T$ corresponding to the energy of the localised states. On the other hand, the behaviour of the decay time with $T$ of the fast component $\tau_2$ suggests that there is a non-radiative contribution even at lowest $T$, which prevents us to use Eq.~(\ref{eq:tau_R_fromtau1}). To overcome this limitation, we assume that the radiative lifetime at 15~K is the same as that for the slow component $\tau_1$,~i.e., $\tau_2^\mathrm{R}(15K)=\tau_1^\mathrm{R}(15K)$, and a $T$ independent non-radiative decay $\tau_\mathrm{C}$ is also present and given by
\begin{eqnarray}
\frac{1}{\tau_\mathrm{C}}=\frac{1}{\tau_2(15K)}-\frac{1}{\tau_2^\mathrm{R}(15K)}\,.\label{eq:tau_C}
\end{eqnarray}

Since $\tau_\mathrm{C}$ is not dependent on $T$, we can now calculate the radiative lifetime $\tau_2^\mathrm{R}$ of the fast component at any $T$ using Eq.~(\ref{eq:tau_R_fromtau1}), replacing $\tau_1$ with $\tau_2$ and $1/\tau_\mathrm{NR}$ with $1/\tau_\mathrm{C}+1/\tau_2^\mathrm{NR}$. The overall decay time as a function of $T$ is then given by
\begin{eqnarray}
\frac{1}{\tau_2(T)}=\frac{1}{\tau_\mathrm{C}}+\frac{1}{\tau_2^\mathrm{R}(T)}+\frac{1}{\tau_2^\mathrm{NR}(T)}\,.
\end{eqnarray}
Hence, we can repeat the analysis of the radiative and non-radiative part described by Eqs.~(\ref{eq:tau_R})--(\ref{eq:nonradiative}) for $\tau_2^\mathrm{R}$ and $\tau_2^\mathrm{NR}$. A similar approach can be used also for $\tau_3$ of samples $\mathrm{S}_\mathrm{with}$ and $\mathrm{S}_\mathrm{cap}$, with the assumption that the same radiative lifetime is used for $\tau_3$ as that for $\tau_2$,~i.e., $\tau_3^\mathrm{R}(15K)=\tau_2^\mathrm{R}(15K)$, and $T$ independent non-radiative lifetime $\tau_\mathrm{C}$ is similar to Eq.~(\ref{eq:tau_C}) as $1 / \tau_\mathrm{C} = 1/ \tau_3(15K)-1/ \tau_3^\mathrm{R}(15K)$. The numerical results of the described deconvolution are summarised in Tab.~\ref{tab:TRPL_temperature} for individual decay times taken at the maximum of the PL intensity for each sample. Based on the previous analysis, we worked out the Arrhenius-like equation with explicit dependence of PL on all parameters derived from the TRPL results:
\begin{widetext}
\begin{equation}
\frac{I_0}{I_\mathrm{PL}(T)}=1+\sum_{i=1}^{2(3)} {\left[\tau_{i\mathrm{R}}^0+\tau_{i\mathrm{R}}^T\exp{\left(\frac{T}{T_{i\mathrm{C}}}\right)}\right] \times \left[\frac{1}{\tau_{i\mathrm{NR}}^1}\exp{\left(\frac{-E_{i1}}{k_\mathrm{B}T}\right)} + \frac{1}{\tau_{i\mathrm{NR}}^2}\exp{\left(\frac{-E_{i2}}{k_\mathrm{B}T}\right)}\right]}, \label{eq:TRPL_Arhenius}
\end{equation}
\end{widetext}
where the upper limit of the sum depends on a number of mono-exponential decays in the fitting model used for deconvolution of the TRPL signal. 

\begin{center}
\begin{table*}[!ht]
{\small
\hfill{}
\caption{Summary of the TRPL Arrhenius-like fits using Eq.~(\ref{eq:TRPL_Arhenius}). The displayed values are obtained with accuracy better than $10^{-2}\%$.}
\begin{ruledtabular}
\begin{tabular}{ccccccccc}
sample&	process & $E_1$ [meV]& $\tau_\mathrm{NR}^1$ [ns]& $E_2$ [meV]& $\tau_\mathrm{NR}^2$ [ns] & $\tau_\mathrm{R}^0$ [ns]& $\tau_\mathrm{R}^T$ [ns]& $T_C$ [K]\\ 	
\hline 
	%\multirow{2}{*}{$\mathrm{S}_\mathrm{w/o}$ }&	 $\tau_1$& 16.6&0.235&441.6 & 0.020& 67.66&0.00&8.21\\
	%&	 $\tau_2$& 16.6&0.235&393.8 & 0.024& 13.27&180&3000\\ \hline
	\multirow{2}{*}{$\mathrm{S}_\mathrm{w/o}$ }&	 $\tau_1$& 16.7&0.234&441.4 & 0.020& 64.62&0.00&8.18\\
	&	 $\tau_2$& 16.7&0.234&339.6 & 0.059&14.66& 0.30&49.3\\ \hline

\multirow{3}{*}{$\mathrm{S}_\mathrm{with}$ }	&	 $\tau_1$&5.2 & 36.18& 64.3& 0.087& 96.45&0.820&21.6\\
	& $\tau_2$&--&--&57.3 & 0.050&  14.61&8.18&62.2\\
		& $\tau_3$&10.0&4.97 & --&--& 8.58&3.13&56.6\\
		 \hline
\multirow{3}{*}{$\mathrm{S}_\mathrm{cap}$ }	&	 $\tau_1$& 23.5&0.237&591.3 & 47.73& 82.62&0&36.9\\ %28.94
	&	 $\tau_2$& 25.4&0.090&284.7 & 0.111& 13.17&1.95&47.1\\ %15.18
	&	 $\tau_3$& 8.1&9.05&-- & --& 3.10&0.106&14.5\\
\end{tabular} \label{tab:TRPL_temperature}
\end{ruledtabular}}
\hfill{}
\end{table*}
\end{center}

We attributed the slowest process $\tau_1$ to the recombination of DAP and other crystalline defects, which follows the same trend with increasing $T$ for $\mathrm{S}_\mathrm{w/o}$ and $\mathrm{S}_\mathrm{cap}$,~i.e., it decreases over 2 orders of magnitude from 100~ns to 1~ns. Due to larger amount of defects, $\tau_1$ of $\mathrm{S}_\mathrm{with}$ decreases only by one order of magnitude to 20~ns, which significantly changes the character of the radiative lifetime, increasing exponentially with $T$ from $\tau_\mathrm{R}^0=96.45$~ns at 15~K due to thermalization of the defects. In comparison with that for $\mathrm{S}_\mathrm{w/o}$ and $\mathrm{S}_\mathrm{cap}$, we find that to be constant at 64.62~ns and 82.62~ns, respectively. 

The radiative time constant $\tau_\mathrm{R}$ of the faster process $\tau_2$ increases exponentially across the samples with $T$ from $\tau_2=$14~ns. This increase is most likely caused by impurity thermalization via $T_\mathrm{C}$ ($T_\mathrm{C}\approx50$\,K is close to disorder energy determined for these samples in~\cite{Steindl2019_PL}). While no material exchange with QD constituents in GaAs IL for sample $\mathrm{S}_\mathrm{w/o}$ occurs by design, confirmed by the fact that the amplitude $\tau_\mathrm{R}^T$ of thermalization change of $\tau_\mathrm{R}$ is almost zero, after QD formation, In-Ga redistribution occurs as previously reported in Refs.~\cite{Steindl2019_PL, Gajjela2020}, leading to almost thirty-fold increase of $\tau_\mathrm{R}^T$ (sample $\mathrm{S}_\mathrm{with}$).
The redistribution can be prevented by overgrowing the structure by a thin GaSb capping layer (see the similarity in panels of $\mathrm{S}_\mathrm{cap}$ and $\mathrm{S}_\mathrm{w/o}$ in Fig.~\ref{fig:Temp_TRPL}), which for a thickness of $\sim$1~ML leads to approximately six-times larger $\tau_\mathrm{R}^T$ than that for sample $\mathrm{S}_\mathrm{cap}$, and an As-Sb intermixing between QDs and capping takes place, resulting in an increase of the Sb content in QDs~\cite{Steindl2019_PL}. 

It can be assumed that the importance of this effect can be reduced if the Sb layer is thicker because then the capping might be more robust, yet that can also result in pushing the wavefunctions out of the QD body, and the corresponding change of the type of spatial band-alignment, previously reported for similar dots grown on GaAs substrate in Refs.~\cite{Klenovsky_IOP2010,Klenovsky2010,Klenovsky2015}. 

The fastest process $\tau_3$ was considered only for QD samples $\mathrm{S}_\mathrm{with}$ and $\mathrm{S}_\mathrm{cap}$. The parameter $\tau_3$ of the sample $\mathrm{S}_\mathrm{with}$ decreases from $\sim 10$~ns (at 15~K) to 6~ns (at 70~K). Since the value of the lifetime is close to $\tau_2$, we assume that the electrons are localized preferably at the QD/IL interface. The radiative part $\tau_\mathrm{R}$ is quenched with $T_\mathrm{C}=56.6$~K, corresponding to thermalization energy of 4.9~meV, which is in good agreement with 4.5~meV, previously extracted from the thermal red shift~\cite{Steindl2019_PL}. The presence of additional Sb during QD formation and ripening, which here would translate into the growth of the GaSb cap right after the QD formation, has very likely led to the formation of smaller and more homogeneous QDs, as a result of the Sb surfactant effect, as also pointed out by Sala~\textit{et al.} in Refs.~\cite{Sala2016, t_sala}. This process could have, thus, led to a better electron-wavefunction localization in the QD body, resulting in a shorter decay time $\tau_3$ of $\approx 3$~ns (at 15~K and decreasing to 2~ns at 70~K) for $\mathrm{S}_\mathrm{cap}$. This is in agreement with the 2.5~ns observed for (InGa)(AsSb)/GaAs/GaP QDs grown with higher Sb flow~\cite{Sala2016}. This points to the fact that both growing a thin GaSb cap above the QDs and using a higher Sb flow before QD formation are both efficient ways to affect the QD structural properties and possibly increase the Sb content in the QDs~\cite{Gajjela2020}.
%increase the Sb content in QDs.
\\The transition is thermally quenched with $T_\mathrm{C}=14.5$~K (1.3~meV is in good agreement with 1.4--2.0~meV extracted from $T$-resolved PL experiments~\cite{Steindl2019_PL}) of $\tau_R$ into disordered centers most likely at the QD/IL interface. The analysis in panels (a) and (b) of Fig.~\ref{fig:Temp_TRPL} shows that PL of the sample $\mathrm{S}_\mathrm{w/o}$ is thermally quenched via phonon-excitation from $X$-valley in GaAs, with activation energy $E_1=16.7$~meV, in good agreement with energies of 10--12~meV extracted from steady-state PL~\cite{Steindl2019_PL}, which was already observed for GaAs/GaP QDs~\cite{t_dagostar}, and for larger $T$ via unipolar escape of electrons from $X$-valley of GaAs layer and GaP to $L$-valley in GaP, with activation energies of $E_2=441.4$~meV (461~meV determined from 8-band $\mathbf{k\cdot p}$) and $E_2=339.6$~meV (370~meV from 8-band $\mathbf{k\cdot p}$)~\cite{Steindl2019_PL}, respectively.

From the analysis of non-radiative lifetime in panels (c)--(e) in Fig.~\ref{fig:Temp_TRPL}, we identify that the emission from sample $\mathrm{S}_\mathrm{with}$ at low $T$ is thermally quenched via electron-thermalization from $X_{xy}$ in IL to, most likely, nitrogen complexes present in the structure from GaP growth~\cite{Skazochkin_GaPtraps}, having escape energies of $8\pm2$~meV, in good agreement with Ref.~\cite{ioffe}. For larger temperatures, the dominant mechanism of quenching with escape energies $\sim$60~meV is most likely the escape of electron from $X_{xy}$-valley in IL to $X$-valley in bulk (41~meV determined from 8-band $\mathbf{k\cdot p}$, $43\pm7$ meV observed in Ref.~\cite{Abramkin2019_GaAsonGaP}). Having lower eigenenergy and many of available electron states, this escape process is preferably comparable to two concurrently possible ones with similar energies -- the escape of electron from $X_{xy}$-valley in IL to $L$-valley in IL (87\,meV) and the escape of $L$-electron in QDs to the bulk GaP (46\,meV).

Also, for the sample $\mathrm{S}_\mathrm{cap}$ we identify, using the same analysis as in panels (f)--(h) of Fig.~\ref{fig:Temp_TRPL}, a shallow impurity ($8.1$~meV), phonon-emission ($\approx 25$~meV), escape of electron from IL to GaP substrate (284.7~meV, from PL 245~meV~\cite{Steindl2019_PL}, 288~meV from 8-band $\mathbf{k\cdot p}$), and hole-escape from IL to bulk ($\approx 590$~meV, 670~meV from 8-band $\mathbf{k\cdot p}$), see Fig.~15 in~\cite{Steindl2019_PL}. Note that we attribute the increase in $E_2$ to correspond to the phonon emission to As-Sb intermixing between GaAs IL and GaSb capping layer, reported already above. Calculating the activation energies by $\mathbf{k\cdot p}$ model,~i.e., without atomistic resolution, cannot explain the observed changes, such as intermixing or material redistribution on the surface of QDs, which creates a concentration gradient leading to local strain and potential changes affecting the escape of carriers and, therefore, a slight discrepancy between experiment and simulation is expected. 

%%%%%%%%%%%%%%%%%%%%%%%%%%%%%%%%%%%%%%%%
% CONCLUSION
%%%%%%%%%%%%%%%%%%%%%%%%%%%%%%%%%%%%%%%%
\section{Conclusions and outlook}
We performed the first detailed analysis of the carrier dynamics of (InGa)(AsSb)/GaAs/GaP QDs to date, by means of energy and temperature modulated time-resolved-photoluminescence. Based on steady-state PL measurements carried out in our previous work~\cite{Steindl2019_PL} as a reference, we develop spectral shape model taking into account phononic, impurity-related, and thermalization effects to address the four emission bands expected from ${\bf k}\cdot{\bf p}$ calculations~\cite{Klenovsky2018_TUB}. The application of analytical models shows similarities across the samples studied here, originating from GaAs interlayer and defects in the GaP substrate. Specifically, the transitions are zero-phonon and phonon-assisted transitions of electrons in the GaAs interlayer from the $X_{xy}$ valley to the $\Gamma$ valence band, with decay times around 15~ns, and donor-acceptor pair recombination in GaP decaying extremely slowly (up to few~$\mu$s). Moreover, we observe type-I emission from QDs, which is faster than 10~ns and its recombination times varies across the studied range, most likely due to coexistence of momentum direct and indirect transitions and compositional changes of individual dots. Finally, we want to point out the spectral shift of the type-I emission from GaAs interlayer and QDs bands caused by charge potentials from defects created during QD formation. This shift is evident in both pump-power resolved photoluminescence, as well as in the time domain study of the emission. %This observation confirms our previous theoretical explanation about the nature of the power-dependent blue-shift of emission from type-II quantum dots~\cite{Klenovsky2017}.

Our data suggest that epitaxial growth strategies can be employed to efficiently increase the Sb content in the QDs by a thin GaSb cap overgrowth. Such Sb concentration increase in QDs increases the carrier confinement and will subsequently lead to an increase of the QD storage time, which is of utmost importance for the implementation of such QDs into nano-memory devices~\cite{Nowozin2013,Bimberg2011_SbQDFlash}. However, the use of Sb, and its potential partial segregation~\cite{Gajjela2020,Desplanque_2017}, may lead to the formation of additional point defects, which could affect the storage time by increasing capture cross-section~\cite{t_nowozin}. Therefore, the development of the truly defect-free Sb-rich QDs on top of GaP is the key for further improvement of QD-Flash nano-memories. In this respect, further epitaxial engineering techniques are demanded. However, considering the present study and our previous work~\cite{Steindl2019_PL}, we have demonstrated that overgrowing such QDs with a GaSb capping layer is a promising epitaxial method to increase the Sb content in (InGa)(AsSb) QDs and to manipulate their carrier dynamics.

Furthermore, for their naturally small FSS~\cite{Klenovsky2018}, such Sb-rich dots are promising candidates for entangled-photon sources, potentially operating not only at cryogenic temperatures due to Sb-increased electron confinement. The use as entangled-photon, as well as single-photon, sources will require future effort in the optimization of optical efficiency by both sample quality and cavity enhancement~\cite{Emberger2013}. Even though the growth may be challenging, these structures have benefits, such as small size and improved compositional homogeneity compared to conventional SK QDs~\cite{Sala2018, t_sala, Gajjela2020}. Moreover, considering the negligible lattice mismatch between GaP and Si, they can serve as a CMOS compatible quantum platform. Finally, since the incorporation of Sb during growth leads to (i) tunable quantum confinement of the dots~\cite{Klenovsky2018_TUB} and (ii) the possibility to reduce the amount of charge trap states originating from crystal structure imperfections, we suppose our dots might be superior to those recently proposed on SiGe quantum dots~\cite{Rauter_ACSPhotonic2018_Ge-DEQD, Murphy2021}.

%%%%%%%%%%%%%%%%%%%%%%%%%%%%%%%%%%%%%%%%
% ACKNOWLEDGEMENTS
%%%%%%%%%%%%%%%%%%%%%%%%%%%%%%%%%%%%%%%%
\section{Acknowledgements}
P.S. is Brno Ph.D. Talent Scholarship Holder--Funded by the Brno City Municipality. 
E.M.S. and D.B. thank the DFG (Contract No. BI284/29-2).
A part of the work was carried out under the project CEITEC 2020 (LQ1601) with financial support from the Ministry of Education, Youth and Sports of the Czech Republic under the National Sustainability Programme II. Project CUSPIDOR has received funding from the QuantERA ERA-NET Cofund in Quantum Technologies implemented within the European Union's Horizon 2020 Programme. In addition, this project has received national funding from the MEYS and funding from European Union's Horizon 2020 (2014-2020) research and innovation framework programme under grant agreement No 731473. The work reported in this paper was (partially) funded by project EMPIR 17FUN06 Siqust. This project has received funding from the EMPIR programme co-financed by the Participating States and from the European Union’s Horizon 2020 research and innovation programme. This works was also partially funded by Spanish MICINN under grant PID2019-106088RB-C3 and by the MSCA-ITN-2020 Funding Scheme from the European Union’s Horizon 2020 programme under Grant agreement ID: 956548.

% \bibliography{paper_TUB_TRPL_QDs.bib}
%%%%%%%%%%%%%%%%%%%%%%%%%%%%%%%%%%%%%%%%
% APPENDIX
%%%%%%%%%%%%%%%%%%%%%%%%%%%%%%%%%%%%%%%%
\pagebreak \renewcommand{\thefigure}{A\arabic{figure}}\setcounter{figure}{0}\renewcommand{\theequation}{A\arabic{equation}}\setcounter{equation}{0}
% \appendix
\section{Appendix}
\subsection{Repumping}
\begin{figure}[!ht]
	\centering
	\includegraphics[width=1\linewidth]{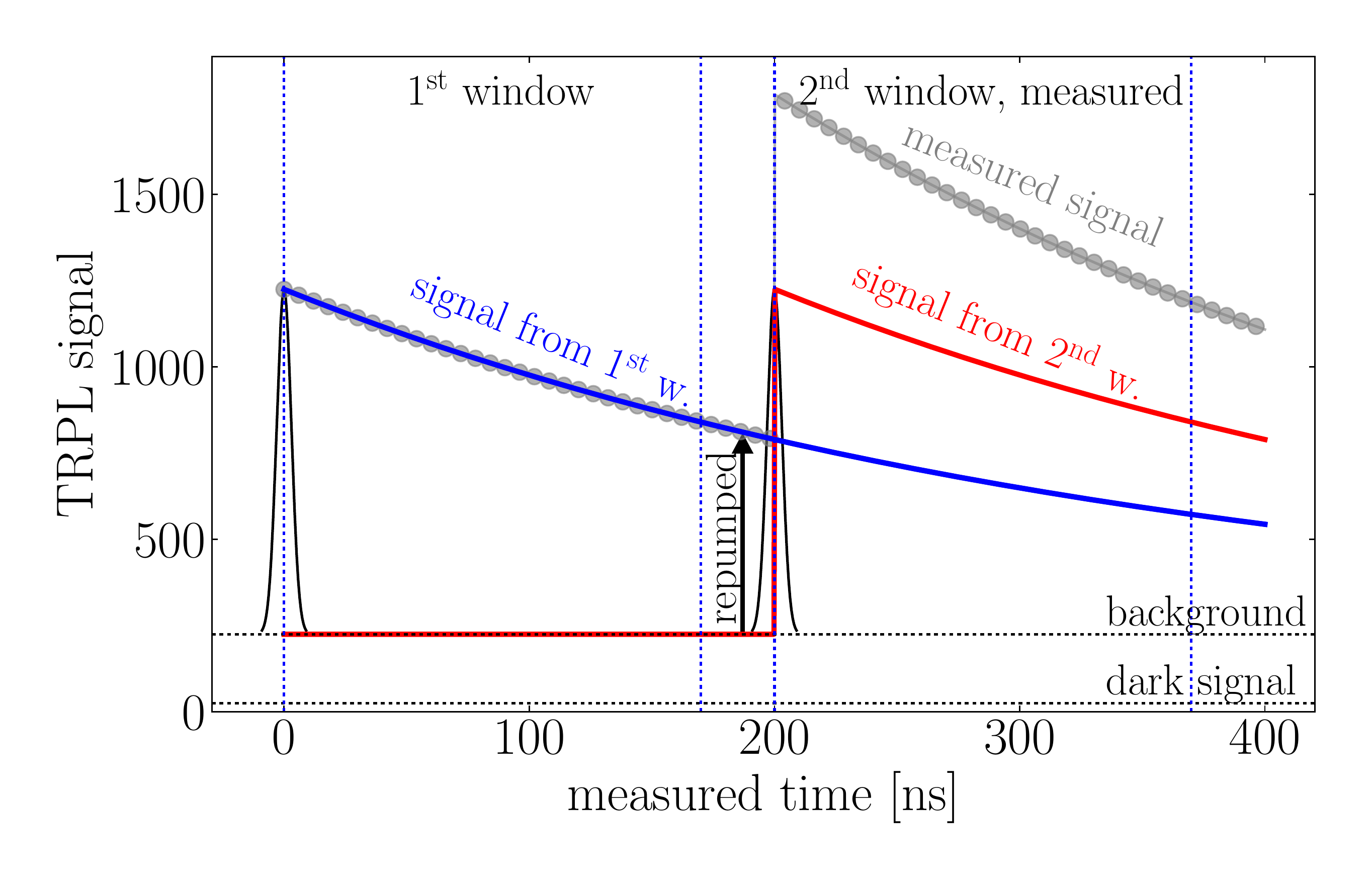}
	\caption{TRPL decay signal with $\tau=350$\,ns (blue for 1$^\mathrm{st}$ window, red for $2^\mathrm{nd}$) after excitation (black) shown in two consecutive temporal windows (200\,ns). Gray symbols represents compound signal from two temporal windows. The arrow points to re-pumped signal from background level (including dark counts) due to contribution to the measured signal from the previous temporal window.}
	\label{fig:Repumping}
\end{figure}
Because some of the observed transitions decay rather slowly and do not completely disappear in one temporal window, we take into account re-pumping of the slow TRPL component $\tau_1$ from previous pulses, which leads to a ``background" increase as can be seen in Fig.~\ref{fig:Repumping}, complicating a proper extraction of the background signal for individual wavelengths and correct time-constant extraction. This issue is overcome in TDPL by disentangling individual transitions by line-shape model fitting, where the slowest decay is assigned to (mainly non-radiative) pair recombination processes of donor-acceptor pairs (DAP) in GaP \cite{Dean_PR68,Dean_1970}.

\end{document}